\documentclass[aps,prb,twocolumn,showpacs,superscriptaddress]{revtex4-1}\begin{tiny}\end{tiny}
\usepackage{longtable}
\usepackage{amsfonts}
\usepackage{graphicx}
\usepackage{setspace}

\begin{document}

\title{Size Effects in Thermal Conduction by Phonons}

\author{Philip B. Allen}
 \email{philip.allen@stonybrook.edu}
 \affiliation{Physics and Astronomy Department, Stony Brook University, Stony Brook, NY 11794-3800, USA}

\date{\today}


\begin{abstract}
Heat transport in nanoscale systems is both hard to measure microscopically, and hard to interpret.
Ballistic and diffusive heat flow coexist, adding confusion.
This paper looks at a very simple case: a nanoscale crystal repeated periodically.  This is a popular model for
simulation of bulk heat transport using classical molecular dynamics (MD), and is related to transient
thermal grating experiments.  Nanoscale effects are seen in perhaps their simplest form. 
The model is solved by an 
extension of standard quasiparticle gas theory of bulk solids.  
Both structure and heat flow are constrained by periodic boundary conditions.   
Diffusive transport is fully included, while ballistic transport by
phonons of long mean free path is diminished in a specific way.
Heat current $J(x)$ and temperature gradient $\nabla T(x^\prime)$ have a
non-local relationship, {\it via} $\kappa(x-x^\prime)$, over a distance $|x-x^\prime|$ determined by phonon mean free paths.  
In MD modeling of bulk conductivity, finite computer resources limit system size.   
Long mean free paths, comparable to the scale
of heating and cooling, cause undesired finite-size effects that have to be removed by extrapolation.
The present model allows this extrapolation to be quantified.
 Calculations based on the Peierls-Boltzmann equation,
using a generalized Debye model, show that extrapolation involves
fractional powers of $1/L$.  It is also argued that heating and cooling should be distributed
sinusoidally (${\dot e}\propto \cos(2\pi x/L)$) to improve convergence of numerics. 
\end{abstract}

\pacs{66.70.-f, 63.22.-m, 65.80.-g}

\maketitle

\section{Introduction}

The linear relation between heat current $\vec{J}$ and temperature gradient $\vec{\nabla} T$ is
\begin{equation}
J_\alpha(\vec{r})=-\int d\vec{r}^\prime \kappa_{\alpha,\beta}(\vec{r},\vec{r}^\prime)\nabla_\beta T(\vec{r}^\prime).
\label{eq:nonlocalr}
\end{equation}
This non-local expression \cite{Mahan} defines the general linear response for a time-independent (steady state) heat flow.
This paper concerns cases where heat flows in a single ($x$) direction in response to a temperature gradient in
the same direction.  Then the spatial variation of $\vec{J}$, $\vec{\nabla}T$, and $\kappa$
can be simplified by averaging over $y$ and $z$ directions.  The conductivity $\kappa_{\alpha,\beta}(\vec{r},\vec{r}^\prime)$
becomes the scalar $\kappa(x,x^\prime)$,
\begin{equation}
J_x(x)=-\int dx^\prime \kappa(x,x^\prime)\nabla_x T(x^\prime).
\label{eq:nonlocal}
\end{equation}
When the distance scale of variation of the temperature gradient
is longer than the carrier mean-free path $\Lambda$,  
a local approximation (the usual Fourier law $J_x=-\kappa_0\nabla_xT$) works.
This is the usual situation
in a macroscopic measurement on a spatially homogeneous sample;
the temperature gradient typically has negligible
spatial variation on the scale of $\Lambda$.  In this limit, $\kappa(x,x^\prime)=\kappa(x-x^\prime)$,
and $\kappa_0 = \int dx \kappa(x)$.  In the homogeneous case,
assuming a long sample, Fourier variables are appropriate.  The 
linear relation is then $J_x(k)=-\kappa(k)\nabla_x T(k)$.  
I will use the same symbol ($\nabla_x T$ for example) to
indicate both coordinate space and reciprocal space representations of functions, the 
coordinate $x$ or $k$ being explicitly shown.  For a homogeneous sample and
constant temperature gradient, the experiment is described by the $k=0$ Fourier
component, and $\kappa_0=\lim_{k\rightarrow0}\kappa(k)$.  

Nanoscale systems have boundaries that add complexity \cite{Dames,Chen,Zhang,Cahill}.
There is an argument \cite{Majumdar} that says the generalized Fourier law 
Eq.(\ref{eq:nonlocalr}) does not apply in all cases.  
This paper avoids such issues, and addresses
a case where the complexity is minimized, namely, a homogeneous nanoscale system
with periodic boundary conditions.  This geometry is experimentally realized in
transient thermal grating experiments \cite{Johnson,Maznev,Collins}.
It is also a preferred geometry for simulation of heat transport 
by classical molecular dynamics (MD) using the ``direct method''
\cite{Visscher,Muller-Plathe,Maiti,Zhou,Cao}.  Heat is introduced locally,
extracted locally some distance away, and temperature is monitored
in between.  The aim of the modeling is usually to extract the conductivity
of a macroscopic sample.  Nanoscales automatically enter, because
the computer cannot process atomic information on a macroscopic scale.

A very nice example was given by Zhou {\it et al.} \cite{Zhou}, who 
carefully analyze computational accuracy, using the semiconductor GaN
as an example.  The large thermal conductivity (several hundred W/mK
at room temperature) indicates that the dominant acoustic phonons have
mean free paths $\Lambda_Q$ of order hundreds of interatomic spacings $a$.
Since acoustic-phonon mean-free paths increase rapidly as wave-vector
$\vec{Q}\rightarrow0$, significant heat is carried by phonons
with much longer mean free paths, thousands in units of $a$.  
The simulations were done for periodic cells up to length $L \approx 1000a$. 
Probably another factor of 10 in length would be required to fully converge the answers.
At a series of lengths $L<1000a$, conductivity $\kappa_{\rm eff}(L)$ was extracted
as the ratio $-J_x/\nabla_xT$, where the temperature gradient was
determined at $x=\pm L/4$, half way between the heat source and sink. 
Extrapolation of $\kappa_{\rm eff}(L)$ to $L\rightarrow\infty$ 
(assuming $\kappa_{\rm eff}(L)-\kappa_{\rm bulk}\propto 1/L$) indicated that 
the value of $\kappa_{\rm bulk}$ was typically twice bigger than the maximum 
achieved value $\kappa_{\rm eff}(L\approx 1000a)$.
Zhou {\it et al.} \cite{Zhou} discover evidence of a breakdown in this method of extrapolation.
The breakdown was analyzed by Sellan {\it et al.} \cite{Sellan}; 
the alternate analysis given here uses similar Debye-model simplifications, but
different boundary assumptions.  Both offer ways of improving extrapolation
needed in MD computation of $\kappa$.

The simplicity of the periodic boundary means that phonon gas theory
can be easily adapted to this nanoscale situation.  Here I offer such an
analysis.  The result supports the idea of Sellan {\it et al.} \cite{Sellan}
that $\kappa_{\rm eff}(L)-\kappa_{\rm bulk}\propto 1/\sqrt L$ 
should give a better extrapolation.  The analysis also points to a better algorithm for
adding and removing heat.

\section{Segmented Periodic Slab Model}

\begin{figure}[top]
\includegraphics[angle=0,width=0.5\textwidth]{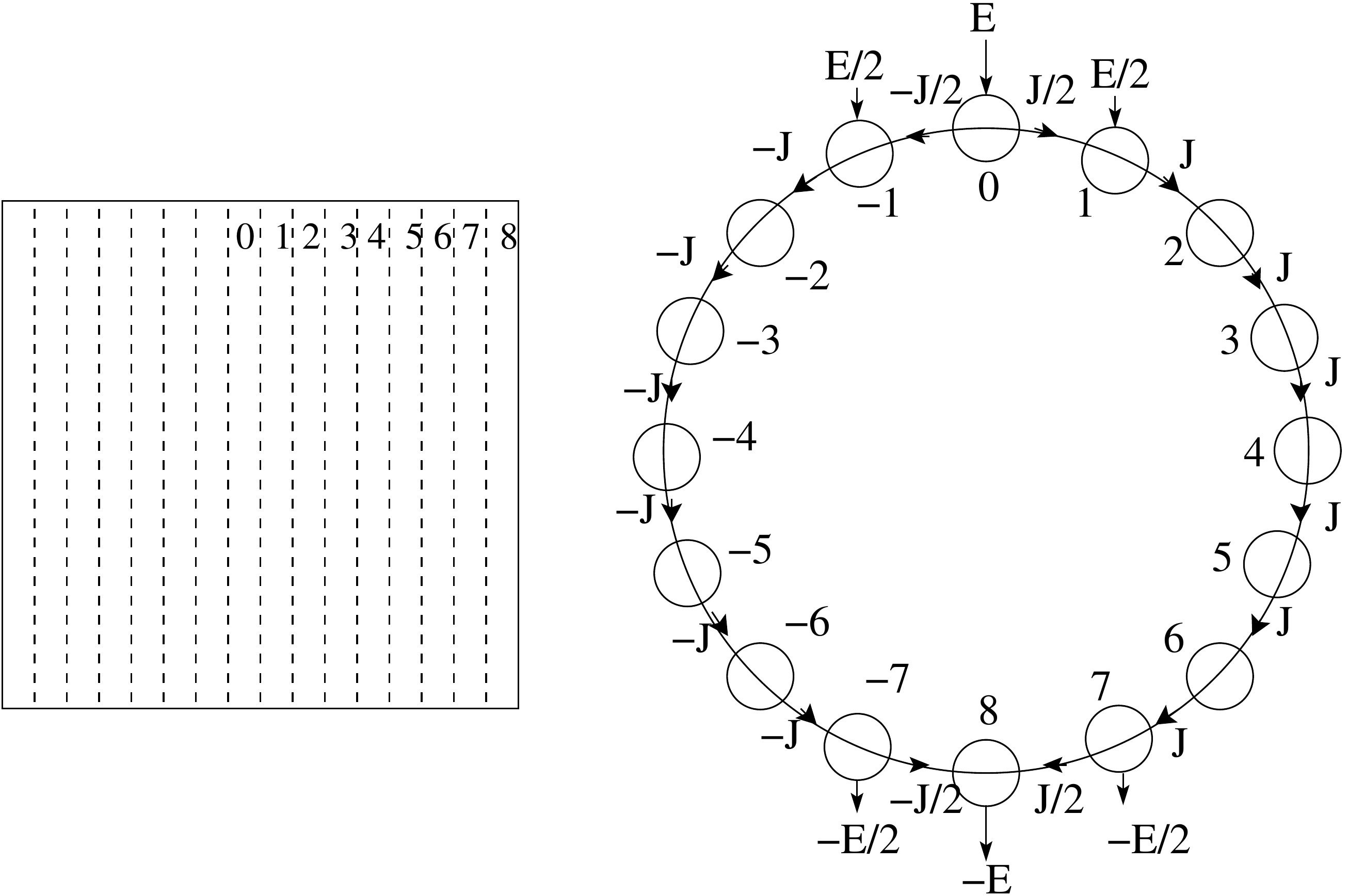}
\caption{\label{fig:ring} Two schematics of a periodic supercell used for MD
modeling of thermal conductivity.  The cell is divided into $N$ slabs (here $N=16$) in the
direction $x$ of heat flow.  It is periodically repeated in all three directions.  
Heat is randomly inserted as impulses on randomly chosen atoms in slab 0.  Equal
random extraction of heat occurs in slab $N/2$.  Thus the heat current density $J$ is controlled.  
Temperature is measured in slabs $1, 2, \ldots, N/2-1$, and also in the similar slabs $-1, -2, \ldots$ .
The temperature gradient $dT/dx$ is minimum in slab $N/4$ and $3N/4$.  From these gradients,
the value of $\kappa_{\rm eff}(N)=J/|dT/dx|_{\rm min}$ is evaluated.}
\end{figure}

Fig. \ref{fig:ring} is a cartoon system (``simulation cell'') of $N$ ``slabs,'' each of width $d$, corresponding
to a total length $L=Nd$.  The system is repeated periodically, as sometimes
used in classical MD simulations.  
The primary variables are the imposed rate of heating per unit volume, ${\dot e}(\ell)$, of the 
$\ell$'th slab, and the ``measured'' slab temperature $T(\ell)$ (mean kinetic energy of the atoms 
in the slab, divided by $3k_B/2$).  
These primary variables coincide 
with the quantities measured in heat conduction experiments.  
The secondary variables are the heat flux $J_x(\ell+\frac{1}{2})$ and the temperature gradient
$\nabla_x T(\ell+\frac{1}{2})$; both are defined at the junction of slabs $\ell$ and $\ell+1$.  
The secondary variables are related to the primary variables by the two
fundamental slab-ring equations,
\begin{equation}
\nabla_x T(\ell+\frac{1}{2})=[T(\ell+1)-T(\ell)]/d
\label{eq:nablaT}
\end{equation}
\begin{equation}
{\dot e}(\ell)=[J_x(\ell+\frac{1}{2})-J_x(\ell-\frac{1}{2})]/d
\label{eq:Jx}
\end{equation}
Equation \ref{eq:Jx} is energy conservation.
To allow a steady state, it is required that $\sum_\ell {\dot e}(\ell)=0$, meaning that no
net heating occurs.

The relation between heat flux
and temperature gradient, in linear approximation, is
\begin{equation}
J_x(\ell+\frac{1}{2})=-\sum_{m=0}^{N-1} \kappa(\ell-m)\nabla_x T(m+\frac{1}{2})
\label{eq:lFou}
\end{equation}
This is the analog of Eq.(\ref{eq:nonlocal}).  
The position variable $x$ has been discretized into the slab index $\ell$.  The thermal
conductivity is similarly discretized.  In terms of these discrete and periodic variables, the form of Eq.(\ref{eq:lFou})
is required by the linear approximation.  The matrix $\kappa(\ell,m)$ is rigorously
defined.  It is independent of the heating ${\dot e}(\ell)$.  If the slabs
are identical, then $\kappa(\ell,m)$ retains
the $N$-fold translational symmetry of the ring.  Therefore $\kappa(\ell+n,m+n)
=\kappa(\ell,m)=\kappa(\ell-m,0)\equiv\kappa(\ell-m)$.
 All variables are $N$-fold periodic in $\ell$, including the
non-local conductivity $\kappa(\ell)=\kappa(\ell+N)$.

\par
\par

Conjugate to the $N$ positions on the ring are $N$ Fourier vectors $q=(2\pi/L)n_q$,
with $L=Nd$ and $n_q$ defined {\it modulo} $N$.  
I will always use $k$ for the Fourier transform of the continuous variable $x$, and
$q$ for the discrete case.  This notation then clarifies that $\kappa(k)$ refers to
thermal conductivity of a large homogeneous system, and $\kappa(q)$ to the
small but periodically repeated supercell.

Just as the integers
$\ell$ are defined {\it modulo} $N$, similarly the integers $n_q$ have the same periodicity,
$n_q+N=n_q$.  The Fourier representation of the discretized variables is particularly
simple and convenient:
\begin{eqnarray}
{\dot e}(\ell)&=&\sum_q e^{iqd\ell}{\dot e}(q) \nonumber \\
T(\ell)&=&\sum_q e^{iqd\ell}T(q) \nonumber \\
J_x(\ell+\frac{1}{2})&=&\sum_q e^{iqd(\ell+\frac{1}{2})}J_x(q) \nonumber \\
\nabla_x T(\ell+\frac{1}{2})&=&\sum_q e^{iqd(\ell+\frac{1}{2})}\nabla_x T(q) \nonumber \\
\kappa(\ell)&=&\sum_q e^{iqd\ell}\kappa(q),
\label{eq:elltoq}
\end{eqnarray}
where sums go over the $N$ distinct values of $q$.  The reverse transforms are
\begin{eqnarray}
{\dot e}(q)&=&\frac{1}{N}\sum_\ell e^{-iqd\ell}{\dot e}(\ell) \nonumber \\
J_x(q)&=&\frac{1}{N}\sum_\ell e^{-iqd(\ell+\frac{1}{2})}J_x(\ell+\frac{1}{2}), 
\label{eq:qtoell}
\end{eqnarray}
and similar for $T$, $\nabla_x T$, and $\kappa$.  
Sums go over the $N$ distinct values of $\ell$.
The reciprocal space version of Eq.(\ref{eq:lFou}) is 
\begin{equation}
J_x(q)=-\kappa(q)\nabla_x T(q)
\label{eq:qFou}
\end{equation}

\section{Phonon Gas Theory}

This section assumes a macroscopic homogeneous (``continuum'') solid.  The next section
translates this to a segmented supercell of periodic (``discrete'') slabs.  The short-hand $Q$,
in both continuum and discrete models, denotes $(\vec{Q},s)$, 
the (three-dimensional) wavevector $\vec{Q}$ and the branch index $s$ of the phonons.  Spatial variation
is driven by external heating and is assumed to occur only in the $x$ direction. 
Thermal properties are described by $N_Q(x)$, the mean occupation of mode $Q$ at (one-dimensional)
position $x$.  The heat current is
\begin{equation}
J_x(x) = \frac{1}{\Omega}\sum_Q \hbar\omega_Q v_{Qx} N_Q(x),
\label{eq:j}
\end{equation}
The Peierls-Boltzmann equation (PBE) \cite{Peierls,Ziman} describes the dynamics of $N_Q(x)$.
Anharmonic phonon events drive $N_Q$ to the local thermal equilibrium Bose-Einstein 
distribution $n_Q = 1/(\exp(\hbar\omega_Q/k_B T(x))-1)$.  Spatial variations
in $N_Q(x,t)$ change by phonon drift with velocity $\vec{v}_Q = \partial \omega_Q/
\partial \vec{Q}$.  In steady state, $N_Q$ is stationary,
\begin{equation}
\frac{\partial N_Q}{\partial t}=0=\left(\frac{\partial N_Q}{\partial t}\right)_{\rm drift}
+\left(\frac{\partial N_Q}{\partial t}\right)_{\rm collisions}.
\label{eq:PBE1}
\end{equation}
Writing $N_Q$ as the sum of a local equilibrium $n_Q (T(x))$ (unaffected
by collisions) and a deviation $\Phi_Q (x) $, the drift term of the PBE has
two contributions, $- (\partial n_Q/\partial T)v_{Qx} \nabla_x T$
and  $-v_{Qx}\nabla_x\Phi_Q$.  The second of these vanishes when the 
temperature gradient is constant, but becomes important when there is spatial
inhomogeneity.  The collision term, after linearization,
has the form $-\sum_{Q^\prime}C_{QQ^\prime}\Phi_{Q^\prime}$, where $C_{QQ^\prime}$
is a complicated collision operator.  To first approximation (the ``relaxation
time approximation,'' RTA) this can be written
as $-\Phi_Q/\tau_Q$, where $1/\tau_Q$ is the ``single mode relaxation rate,''
meaning the thermalization rate (or life-time broadening) of mode $Q$ that 
appears if only that one mode is out of  equilibrium.  The PBE then takes the form
\begin{equation}
[v_{Qx}\nabla_x + 1/\tau_Q]\Phi_Q(x)=-v_{Qx}\frac{\partial n_Q}{\partial T} \nabla_x T
\label{eq:PBE2}
\end{equation}
In the continuum picture, the direct and reciprocal-space representations are related by 
$\Phi_Q(x)=(1/2\pi)\int dk \exp(ikx)\Phi_Q(k)$ and 
$\Phi_Q(k)=\int dx \exp(-ikx)\Phi_Q(x)$.  The PBE becomes
\begin{equation}
(ikv_{Qx} +1/\tau_Q)\Phi_Q(k)=-v_{Qx}\frac{\partial n_Q}{\partial T}\nabla_x T(k).
\label{eq:PBE3}
\end{equation}
The thermal conductivity $\kappa(k)$ in continuous $k$ space obeys the
reciprocal-space version of Eq.(\ref{eq:nonlocal}),
$J_x(k)=-\kappa(k)\nabla_x T(k)$, and has the form
\begin{equation}
\kappa(k)=\frac{1}{\Omega} \sum_Q \frac{\hbar \omega_Q v_{Qx}^2 (\partial n_Q/\partial T)}
{1/\tau_Q + ikv_{Qx}}
\label{eq:kcq}
\end{equation}
This can be Fourier-transformed back to $x$-space:
\begin{equation}
\kappa(x-x^\prime)=\frac{1}{\Omega}\sum_Q^{v_{Qx}>0}
\hbar\omega_Q v_{Qx} \frac{\partial n_Q}{\partial T}
\exp \left[-\frac{|x-x^\prime|}{\Lambda_{Qx}}\right],
\label{eq:kcx}
\end{equation}
where the mean free path is $\Lambda_{Qx} = v_{Qx}\tau_Q$.
This uses the property that $v_{Qx}$ and $\Lambda_{Qx}$ change sign when
$\vec{Q}$ changes sign.  

The result Eq.(\ref{eq:kcx}) shows explicitly that heat transport is
influenced non-locally by temperature variations.  The velocity
$|v_{Qx}|$ is bounded, but the mean free path $\Lambda_{Qx}=v_{Qx}\tau_Q$ is not.
Relaxation times of long wavelength acoustic phonons diverge
as $|\vec{Q}|\rightarrow 0$.  In the over-simplified (``gray'')
model where every phonon $Q$ has the same relaxation rate $1/\tau_Q=1/\tau_0$,
the $Q$-integrated $\kappa(x)$ decays rapidly.  But because the spatial
decay rate in Eq.(\ref{eq:kcx}) involves $1/\Lambda_0 |\cos\theta_Q|$, no simple
exponential formula works.  The decay is more rapid than the simple  exponential $\exp(-|x|/\Lambda_0)$.
However, in more realistic models, the divergence of $\tau_Q$ 
and $\Lambda_{Qx}$ at small $Q$ causes, after $Q$-integration, a slower than exponential decay.
Still, in Eq.(\ref{eq:kcx}), even long-wavelength phonons are diffusive because the sample
is ``macroscopic.''

\section{Gas Theory on the Segmented Periodic Slab}

%

Gas theory translates to the segmented supercell in a slightly awkward way.
Temperature $T(\ell)$ is a slab property, but current $J_x(\ell+\frac{1}{2})$
is a junction property.  Gas theory has temperature as a primary variable,
so $T(\ell)$ should correspond to $T(x)$ averaged over the $\ell$'th slab.
Evidently $N_Q(\ell)$ is a slab property, not a junction property.  But
since the current in gas theory is $\sum_Q \hbar\omega_Q v_{Qx} N_Q/\Omega$,
we are forced to compromise and define
\begin{equation}
J_x(\ell+\frac{1}{2})=\frac{1}{\Omega}\sum_Q \hbar\omega_Q v_{Qx}
\frac{N_Q(\ell)+N_Q(\ell+1)}{2}.
\label{eq:scurrent}
\end{equation}
Transforming to the discrete slab Fourier representation, this becomes
\begin{equation}
J_x(q)=\frac{1}{\Omega}\sum_Q \hbar\omega_Q v_{Qx}\cos(qd/2)\Phi_Q(q).
\label{eq:Jsq}
\end{equation}
Similarly, the temperature gradient, Eq.(\ref{eq:nablaT}), in Fourier variables, is
\begin{equation}
\nabla_x T(q)=2i\sin(qd/2)T(q)/d,
\label{eq:gttq}
\end{equation}
and the fundamental connection Eq.(\ref{eq:Jx}) between heat input and
current is
\begin{equation}
{\dot e}(q)=2i\sin(qd/2)J_x (q)/d.
\label{eq:dotejq}
\end{equation}

The remaining task is to translate Eq.(\ref{eq:kcq}) to slab language.  The ingredient needing
translation is  $\nabla_x$ which appears twice in Eq.(\ref{eq:PBE2}).  The translation of
this equation is
\begin{eqnarray}
\Phi_Q(\ell)&=&\theta(v_{Qx})v_{Qx}\tau_Q /d \nonumber \\
&\times&\left[ \frac{\partial n_Q}{\partial T}(T(\ell-1)-T(\ell))+(\Phi_Q(\ell-1)-\Phi_Q(\ell))\right] \nonumber \\
&+&\theta(-v_{Qx})v_{Qx}\tau_Q /d \nonumber \\
&\times&\left[ \frac{\partial n_Q}{\partial T}(T(\ell)-T(\ell+1))+(\Phi_Q(\ell)-\Phi_Q(\ell+1))\right], \nonumber \\
&&
\label{eq:PBEd1}
\end{eqnarray}
where $\theta(x)$ is the unit step function.
The meaning is that drift entering slab $\ell$ from the left uses positive velocity phonons which carry
information from the slab on the left, while drift entering slab $\ell$ from the right uses negative
velocity phonons bringing information from the slab on the right.

The discretized PBE, Eq.(\ref{eq:PBEd1}) can be solved in discrete reciprocal space, giving
\begin{equation}
\Phi_Q(q)=-\frac{v_{Qx}\tau_Q(\partial n_Q/\partial T)\nabla_x T(q)}
{S(q)+2i\sin(qd/2)v_{Qx}\tau_Q/d} 
\label{eq:sPBEd}
\end{equation}
where $S(q)$ is $\exp(iqd/2)$ if $v_{Qx}$ is positive
and $\exp(-iqd/2)$ if $v_{Qx}$ is negative.  This gives the answer for the discrete slab gas theory
thermal conductivity,
\begin{eqnarray}
\kappa(q)&=&\frac{1}{\Omega}\sum_Q \hbar\omega_Q\frac{\partial n_Q}{\partial T} 
v_{Qx}^2 \tau_Q \cos(qd/2)  \nonumber \\
&\times&  \left[ \frac{1}{S(q)+2i\sin(qd/2)v_{Qx}\tau_Q /d}  \right]
\label{eq:kappaslab}
\end{eqnarray}
This agrees with the continuum formula Eq.(\ref{eq:kcq}) if the small
q limit is taken, $\exp(iqd/2)\approx 1\approx \cos(qd/2)$ and $2\sin(qd/2)\approx qd$.
Since $v_{Qx}$ changes sign and $S(q)$ becomes $S(q)^\ast$ when $\vec{Q}$ goes to $-\vec{Q}$,
the sum in Eq.(\ref{eq:kappaslab}) is real.  Taking the real part of
the factor in brackets, Eq.(\ref{eq:kappaslab}) becomes
\begin{equation}
\kappa(q)=\frac{1}{\Omega}\sum_Q \hbar\omega_Q\frac{\partial n_Q}{\partial T} 
v_{Qx}^2 \tau_Q \cos^2(qd/2)   F(q,\Lambda_{Qx})
\label{eq:kappaslab1}
\end{equation}
\begin{equation}
F(q,\Lambda)=  \left[ 1 +4\sin^2 (qd/2)
\left\{ \left( \frac{\Lambda}{d} \right) + \left( \frac{\Lambda}{d} \right)^2 \right\} \right]^{-1} 
\label{eq:kappaslab2}
\end{equation}
where $\Lambda_{Qx}=v_{Qx}\tau_Q$.
Now consider what happens at the smallest $q$, namely $q_{\rm min}=2\pi/L$.
The factor $\cos^2(q_{\rm min}d/2)=\cos^2(\pi/N)$ can be set to 1 and $\sin^2(q_{\rm min}d/2)$
can be set to $(q_{\rm min}d/2)^2$ when
the number $N$ of slabs is large.  The correction factor in  Eq.(\ref{eq:kappaslab2})
becomes, in the large $N$ case,
\begin{equation}
F(q_{\rm min},\Lambda_{Qx})\approx \left[ 1+\left( \frac{2\pi\Lambda_{Qx}}{L} \right)^2 \right]^{-1}
\label{eq:corr}
\end{equation}
The part of Eq.(\ref{eq:kappaslab2}) linear in $\Lambda/d$ is neglected compared to the quadratic
part since the correction is only important when $\Lambda/d\gg 1$.  Equations \ref{eq:kappaslab1}
and \ref{eq:corr} show that when a mean free path $|\Lambda_{Qx}|$ becomes comparable
to $L/2\pi$, the contribution of that phonon to $\kappa(q_{\rm min})$ starts to be suppressed,
the suppression becoming complete for phonons $Q$ with $2\pi\Lambda_{Qx} \gg L$.  The reason is,
if a phonon's mean free path is as large as the period (slab ring circumference), that phonon is 
now carrying heat from hotter regions to random regions (after cycling around the ring multiple times.)
This is a different version of a well-known phenomenon in mesoscale heat transport, where $\kappa$
in a sample of size $L$ is diminished because long wavelength phonons travel ballistically a shorter distance $L$, rather than
diffusing the longer distance $\Lambda$ that they would exhibit in bulk \cite{Minnich,Hua}.
Computed behavior based on Eqs. \ref{eq:kappaslab1} and \ref{eq:kappaslab2} 
will be shown in the next section, using a Debye model.

The results developed above enable predictions of $\nabla_x T(\ell+\frac{1}{2})$
and $T(\ell)$ for any given input ${\dot e}(\ell)$.  The idea is to use Eqs.(\ref{eq:elltoq},
\ref{eq:qFou}) to give
\begin{eqnarray}
\nabla_x T(\ell+\frac{1}{2})&=&-\sum_{q\ne 0} e^{iqd(\ell+\frac{1}{2})}J_x(q)/\kappa(q) \nonumber \\
&=& -\sum_{q\ne 0} e^{iqd\ell} \frac{{\dot e}(q)d}{(1-e^{-iqd})\kappa(q)},
\label{eq:dt}
\end{eqnarray}
where the second line follows from the first by using Eq.(\ref{eq:dotejq}).
This can be evaluated using Eq.(\ref{eq:kappaslab}) for $\kappa(q)$.  Using
Eq.(\ref{eq:gttq}), the temperature $T(\ell)$ can also be evaluated, from
\begin{equation}
T(\ell)=T_0 + \sum_{q\ne 0} e^{iqd\ell} \frac{{\dot e}(q)d^2}{4 \sin^2(qd/2) \kappa(q)},
\label{eq:t}
\end{equation}
where $T_0$ is the average temperature.  The $q=0$ term of these sums is omitted
because Eqs.(\ref{eq:nablaT},\ref{eq:Jx}) make it clear that $\nabla_x T(q=0)=\sum_\ell \nabla_x T(\ell) =0$,
and similarly for ${\dot e}(q=0)$.

Two approximations have been made.  One is the RTA.  The other 
is discretization error.  Temperature is a statistical variable,
not definable except by averaging over a finite volume.  Therefore, discretization
over a small width $d$ should not cause noticeable error.  However, a problem arises
because gas theory has been forced to conform to the
slab/junction dichotomy of the discrete picture.  This causes the current in 
the $\ell+\frac{1}{2}$ junction to be tied to the temperature both of the 
two slabs $\ell+1$ and $\ell+2$ to the right, minus the temperature of both of the two
slabs $\ell$ and $\ell-1$ to the left.  This should not be a problem for gas theory,
since the theory requires mean free paths longer than the small atomic dimensions
used for slab widths.  Only if mean free paths are shorter than interatomic spacings
(that is, a liquid rather than gas limit) is the current unaware of the temperature beyond 
the two slabs adjacent to the junction.  
Nevertheless, a problem arises when the heat input ${\dot e}(\ell)$ is confined
to a single site ($\ell=0$ for example.)  Then discretization introduces singular
responses in the form of absurd oscillations in the (unphysical) limit of very short
mean free path.  This problem can be cured by distributing the heat input over three slabs.
Specifically, the cure is to use
${\dot e}(\ell)={\dot e}/2$ when $\ell=0$, and ${\dot e}/4$ when $\ell=\pm 1$.
This modification has more ``realism'' than a single-site input.  Still, it is surprising that
it is required in order for gas theory to work smoothly in a slab model.
Within the homogeneous slab model, the discrete non-local conductivity (Eq. \ref{eq:lFou}) is an
exactly defined concept, as is its Fourier representation $\kappa(q)$.  Equations (\ref{eq:dt},\ref{eq:t})
are exact connections within linear response, while Eq. \ref{eq:kappaslab} uses the 
PBE, plus a further (and not essential) simplification, the RTA.

\section{Debye Model}

Full solution of the PBE,
using accurate phonon properties from density functional theory (DFT), is now widely
available \cite{Li}.  Nevertheless, it is useful to have a simplified model as a standard to
compare real calculations against.  The Debye model replaces the Brillouin zone by a sphere
of radius $Q_D=(6\pi^2 n)^{1/3}$, where $n$ is the number of atoms per unit volume.
There are three branches of phonons, approximated by $\omega_Q=v|\vec{Q}|$ with the
velocity $v$ the same for each branch, and three orthogonal directions $\vec{v}_Q$.  The maximum
frequency phonon ($\omega_{Q,{\rm max}}=\omega_D=vQ_D$) occurs at the edge of the 
Debye sphere where $|\vec{Q}|=Q_D$.  
To establish a notation, the specific heat can be written as 
\begin{equation}
C(T)=C_\infty \int_0^{\omega_D} d\omega c(\omega)
\label{eq:C}
\end{equation}
where $c(\omega)={\cal D}(\omega) E(\omega)$, and $C_\infty=3nk_B$
is the high $T$ classical value of $C(T)$.  The factor ${\cal D}(\omega)$ is
the phonon density of states, normalized to 1.  In Debye approximation, this is
 ${\cal D_D}(\omega)=3\omega^2/\omega_D^3$.  The other factor is
 $E(\omega)=[(\hbar\omega/2k_B T)/\sinh(\hbar\omega/2k_B T)]^2$, the Einstein
 formula for the specific heat (in units $k_B$) of one vibrational mode.  This is replaced by
 the classical limit $E=1$ ($C(T)=C_\infty$)
when comparing with classical MD results.
 
In the spirit of the Debye model, the mean free path $\Lambda_Q=v\tau_Q$ 
can be modeled as $\Lambda/\Lambda_{\rm min}(T)=(Q_D/|\vec{Q}|)^p$, or
equivalently, $(\omega_D/\omega)^p$.  The
exponent $p$ depends on details of scattering.  For point impurities such as isotopic
substitutions, $p$ takes the Rayleigh value $p=4$.  
For ``Normal'' (N, not Umklapp, or U) anharmonic scattering, Herring  \cite{Herring}
found $p=1$ for transverse and $p=2$ for longitudinal acoustic modes at low $\omega$.  
For anharmonic U scattering, which is more relevant here, it is usually
argued that $p=2$, but both $p=3$ and
$p=4$ have some support from numerical calculations \cite{Broido,Esfarjani}. 
The minimum mean free path, $\Lambda_{\rm min}$ is found at $|\vec{Q}|=Q_D$.
It has a temperature-dependent value, scaling as $1/T$ from anharmonic scattering in the
classical high $T$ limit.  

\begin{figure}[top]
\includegraphics[angle=0,width=0.5\textwidth]{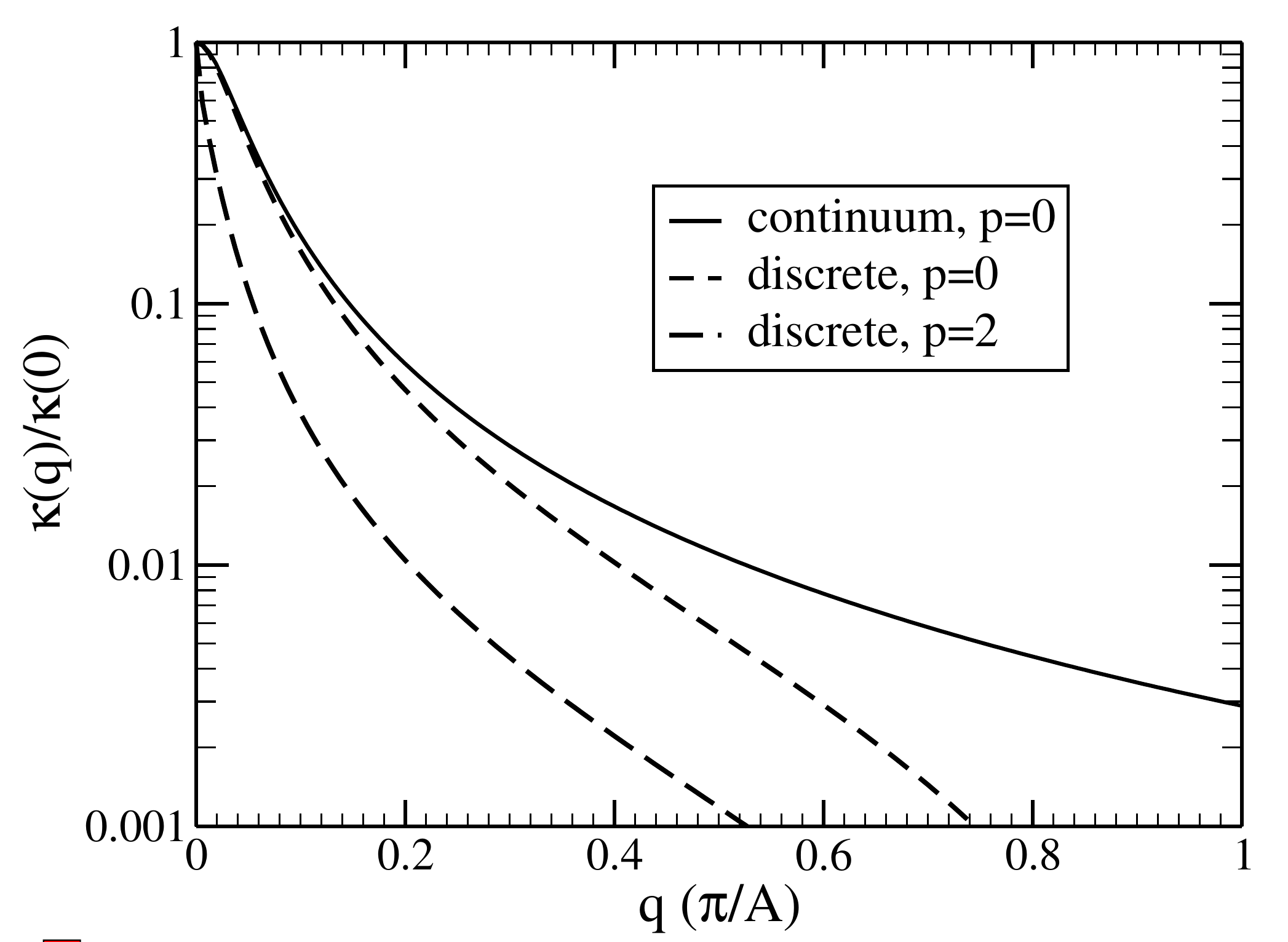}
\caption{\label{fig:kappa} Thermal conductivity {\it versus} wavevector, computed in 
Debye approximation for $\Lambda_{\rm min}=10d$, and plotted versus wavevector
$0\le q \le \pi/d$.  The top curve is the continuum model, Eqs.(\ref{eq:kDcont},\ref{eq:gofu}) 
with constant mean free path $\Lambda=\Lambda_{\rm min}$.  The middle curve is for the discrete
model, Eq.(\ref{eq:kDslab}) with constant mean free path $p=0$.  The bottom curve, from
Eq.(\ref{eq:kDslabp}), uses a more realistic frequency-dependent mean free path, with $p=2$.}
\end{figure}

First consider the continuum theory.  The Debye version has a simple answer
in the unphysical case of $p=0$, where all phonons have the same mean free path, $\Lambda_{\rm min}$.
From Eq.(\ref{eq:kcq}), the answer is
\begin{equation}
\kappa_{D,p=0}(k)=\kappa_0(T)  g(k\Lambda_{\rm min}),
\label{eq:kDcont}
\end{equation}
\begin{equation}
\kappa_{D,p=0}(k\rightarrow 0)\equiv \kappa_0(T)=\frac{1}{3}C(T) v\Lambda_{\rm min}(T).
\label{eq:kap1}
\end{equation}
The function $g(k\Lambda_{\rm min})=g(u)$ is defined as
\begin{equation}
g(u)=\frac{3}{2}\int_{-1}^1 d\cos\theta \frac{\cos^2\theta}{1+iu\cos\theta}=\frac{3(u-\tan^{-1}u)}{u^3}.
\label{eq:gofu}
\end{equation}
The variable of integration, $\cos\theta$, is the cosine of the
angle between the (3d) phonon wavevector $\vec{Q}$ and the
direction ${\hat x}$ of the applied temperature gradient.
The function $g(u)=\kappa_{D,p=0}(k)/\kappa_0(T)$ 
is plotted in Fig. \ref{fig:kappa} for the case where $\Lambda_{\rm min}=10d$. 
In the small $k$ limit, the value is $g\approx 1-3u^2 /5$.

Eq.(\ref{eq:kDcont}) for the more physical case of $p>0$ becomes
\begin{equation}
\frac{\kappa_{D,p}(k)}{\kappa_\infty}= \int_0^{\omega_D}d\omega c(\omega) \left(\frac{\omega_D}{\omega}\right)^p 
g\left(k\Lambda_{\rm min}\left(\frac{\omega_D}{\omega}\right)^p\right),
\label{eq:kDcontp}
\end{equation}
where $\kappa_\infty=C_\infty v\Lambda_{\rm min}(T)/3$ is a convenient scale factor.

For exponent $p\ge 3$, Eq.(\ref{eq:kDcontp})  diverges in the $k\rightarrow 0$ limit.
The low-$\omega$ limit diverges logarithmically at $p=3$, and more severely at
higher $p$.  In theory, the divergence is cut off by finite sample size.  This is rarely seen
experimentally, since $p=2$ ``N scattering,'' 
and Akhieser damping \cite{Akhieser,Maris} provide alternatives.
Finite sample size introduces a complication.  The boundaries destroy the homogeneity
of the theory.  Even phonons with $k$ significantly larger than $k_{\rm min}$ have $\Lambda>L$,
and carry heat ballistically, unaware of the spatial variation of sample temperature $T(x)$.  
To first approximation, this gives a lower limit ($\omega(\Lambda=L)$)
below which the integral Eq.(\ref{eq:kDcontp}) is cut off.  Phonons with
$\omega_{\rm min} < \omega < \omega(\Lambda=L)$ give ballistic currents, less than
their bulk diffusive contribution, and outside the usual local version of 
the Fourier law.  In bulk crystals, this contribution can be important at low $T$, but not at higher $T$
where such phonons are a very small minority and the ballistic component is negligible.

\par
\par

The Debye model can also be used to find expressions for the non-local conductivity
$\kappa(q)$ of the periodic slab model, analogous to $\kappa(k)$ of the continuum model.
Note that $q$ and $k$ (discrete and continuous wavevector) are the notational clue
indicating the model being solved.  First take the unphysical model with
constant mean free path ($p=0$).  Using Eq.(\ref{eq:kappaslab}), the analog of Eq.(\ref{eq:kDcont}) is
\begin{equation}
\kappa_{D,p=0}(q)=\kappa_0(T) h^\prime(q,\Lambda_{\rm min})=\kappa_0(T) h(w)
\label{eq:kDslab}
\end{equation}
where $h^\prime(q,\Lambda_{\rm min})=h(w)$ is
\begin{eqnarray}
h(w)&=&\frac{3\cos(qd/2)}{2}\left[\int_0^1 d\cos\theta\frac{\cos^2 \theta}
{e^{iqd/2}+iw\cos\theta}\right. \nonumber \\
&+&\left. \int_{-1}^0 d\cos\theta\frac{\cos^2 \theta}{e^{-iqd/2}+iw\cos\theta}\right].
\label{eq:h}
\end{eqnarray}
Here $w=2\sin(qd/2)\Lambda_{\rm min}/d$ is the discretized version of $u=k\Lambda_{\rm min}$
that appears in Eq.(\ref{eq:gofu}).  Note that, although $h$ depends separately on $q$ and on
$\Lambda_{\rm min}$, the additional $q$-dependence beyond that contained in $w$ plays only the
role of a fixed parameter, while the $w$ dependence acquires additional importance when the
mean-free path $\Lambda_Q$ acquires $\omega_Q$-dependence.  

Performing the $d\cos\theta$ integral gives
\begin{equation}
h(w)=\frac{3\cos(qd/2)}{w^3}{\rm Re}\left\{ w e^{iqd/2}+ie^{iqd}\ln\left[1+iw e^{-iq/d2}\right]\right\}.
\label{eq:hofw}
\end{equation}
This reduces to Eq.(\ref{eq:gofu}) in the small $q$ limit, under the replacements 
$q\rightarrow k$ and $w\rightarrow u$.
The function $h=\kappa_{D,p=0}(q)/\kappa_0(T)$ is also shown in Fig. \ref{fig:kappa}.  Up until
$q\approx 0.2\pi/d$ ($q\approx 2\pi/\Lambda_{\rm min}$) the discrete and continuum versions fall almost equally
rapidly with $q$ to $< 0.1$.  Beyond, the discrete case falls increasingly rapidly, going to 0 at
the zone boundary, $q=\pm \pi/d$.  This means, for example, that there is no response to 
input heating ${\dot e}(\ell)
= {\dot e}\exp(\pm i\pi\ell)=(-1)^\ell{\dot e}$.  The reason is that adjacent junctions have 
currents $J(\ell+\frac{1}{2})=\pm {\dot e}/2$.  Therefore, the slab current (the average of the
two adjacent junction currents) is zero.

\begin{figure}[top]
\includegraphics[angle=0,width=0.5\textwidth]{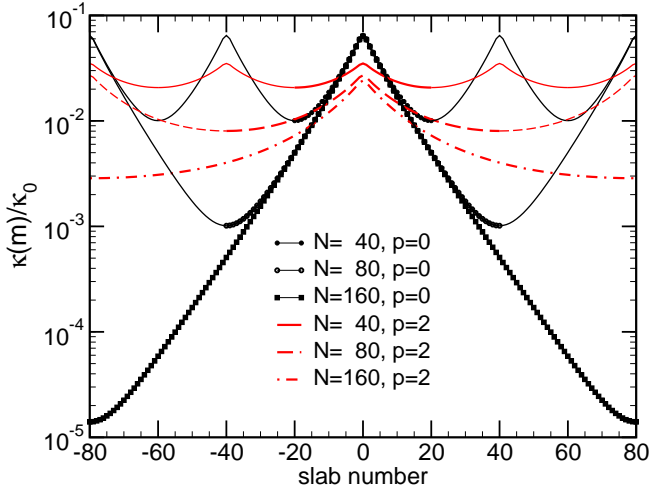}
\caption{\label{fig:kapreal} The non-local thermal conductivity, $\kappa(m)$ defined in Eq.(\ref{eq:lFou}), and
computed using Eq.(\ref{eq:kDslabp}) for the slab model, Fourier transformed back to coordinate space.
The minimum mean free path $\Lambda_{\rm min}$ is $10d$; the supercell has $N$ slabs of width $d$,
where $N$ is shown in the figure legend.  The period $N$ repetitions are shown.
Black and red curves use the power law $p=0,2$ respectively of $\Lambda_Q=
\Lambda_{\rm min}(\omega_D/\omega_Q)^p$.}
\end{figure}

For the more physical case of $p>0$, the answer is
\begin{equation}
\frac{\kappa_{D,p}(q)}{\kappa_\infty}= \int_0^{\omega_D}d\omega c(\omega) \left(\frac{\omega_D}{\omega}\right)^p 
h\left(q\Lambda_{\rm min}\left(\frac{\omega_D}{\omega}\right)^p\right)
\label{eq:kDslabp}
\end{equation}
The function $\kappa_{D,p=2}(q)/\kappa_\infty$ is shown in the classical limit ($c(\omega)={\cal D}(\omega)$)
in Fig. \ref{fig:kappa} as the bottom curve.  The conductivity falls more much rapidly with $q$,
which means increased non-locality.  
This is not surprising.  Spatial memory extends much farther because of the longer mean free paths.
These results are translated back to coordinate space in Fig. \ref{fig:kapreal}.  The $p=0$ (constant
$\Lambda$) results fall exponentially (as $\exp(-md/\Lambda)$) if the supercell size $Nd$ exceeds $\Lambda$ sufficiently.
The $p=2$ case behaves differently.  At a distance $md=8\Lambda_{\rm min}=80d$, the
value of $\kappa(m)$ is $\approx \kappa(0)/8.4$, falling much more slowly than an exponential.
%
\par
\par

\section{Numerical Results}

\par
\begin{figure}[top]
\includegraphics[angle=0,width=0.5\textwidth]{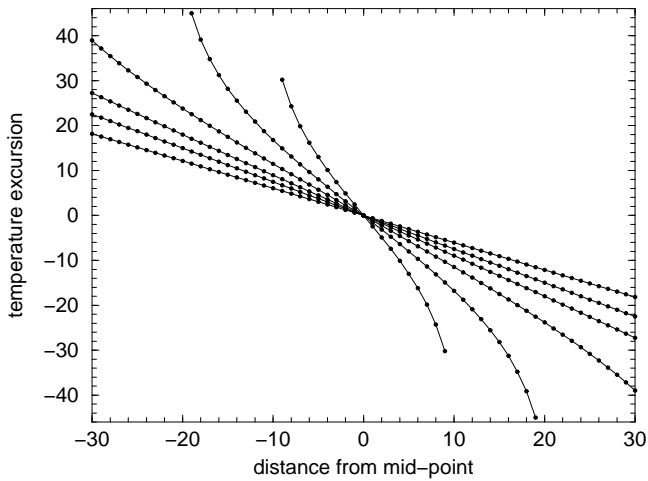}
\caption{\label{fig:temp} Temperature {\it versus} distance from the midpoint
$\ell=N/4$ between hot ($\ell=0$) and cold ($\ell=N/2$) points.  All computations
used the discrete slab formula (Eq.(\ref{eq:kDslabp})) and $\Lambda_{\rm min}=20d$. 
The power law $1/\tau_Q$ is $\omega_Q^2$.  The cell size $N$ was 40 for the
steepest curve, then 80, 160, 320, 640, and 2560.}
\end{figure}
\par

Fig. \ref{fig:temp} illustrates the computed spatial temperature variation for the discrete slab model.
The behavior closely resembles that found by Zhou {\it et al.} \cite{Zhou} in their classical MD simulation
of GaN.  Therefore, I believe that the PBE, as extended here to discrete slabs, and modeled in
RTA and in the Debye approximation, correctly captures the physics.
The calculations of Fig. \ref{fig:temp}  use the classical limit $C(T)=C_\infty$, with mean free path 
$\Lambda=\Lambda_{\rm min}(\omega_D/\omega)^2$, $\Lambda_{\rm min} = 20d$,
and various total cell lengths $L$ ranging from $40d$ to $2560d$.  Symmetry requires 
$T(N/4+m)=-T(N/4-m)$, and an inflection point in $T(\ell)$ at $\ell=N/4$. At the largest
$N$ shown, the distance $N/4$ between heat input and sink is $32\Lambda_{\rm min}$,
and the answer for $\kappa_{\rm eff}(N)$ is still 10\% lower than the macroscopic 
($N\rightarrow\infty$) limit.  The temperature profile is accurately linear, over the 60 slabs shown,
for the three largest lengths $L$, but the slopes are 10 to 30\% higher than the macroscopic limit.

\par
\begin{figure}[top]
\includegraphics[angle=0,width=0.5\textwidth]{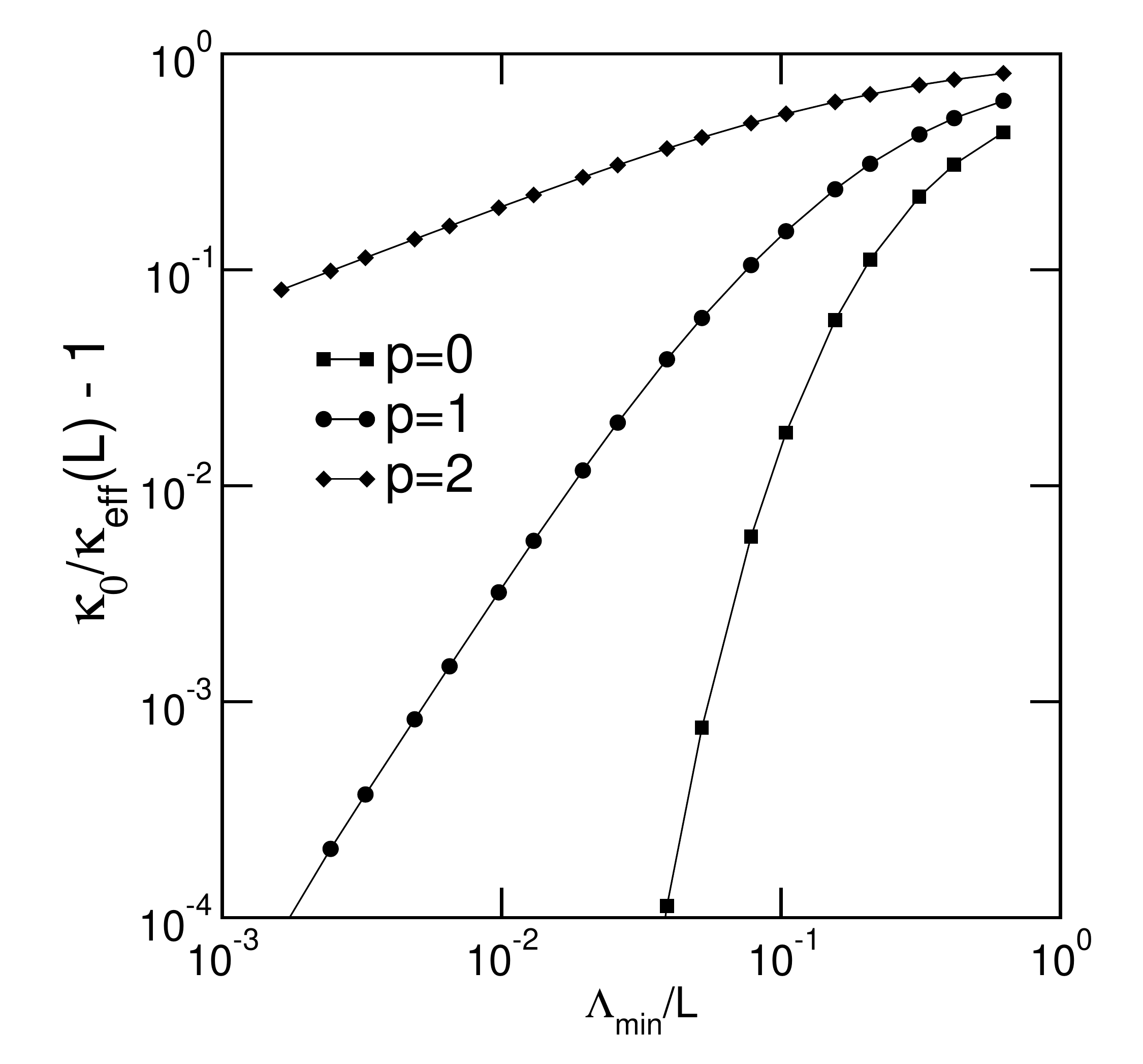}
\caption{\label{fig:power} Rate of convergence of $1/\kappa_{\rm eff}(L)$ to its $L\rightarrow\infty$
limit, by a logarithmic plot as a function of $\Lambda_{min}/L$.  All computations
used the discrete slab formula (Eq.(\ref{eq:kDslabp})) and $\Lambda_{\rm min}=10$. 
From bottom to top, the power law $p$ of $1/\tau_Q \propto \omega_Q^p$ is
$p=$ 0, 1, and 2.  The numerical power law of $\kappa_0/\kappa -1 \propto (1/L)^s$ is $s\approx 
\infty$, 2, and 1/2.}
\end{figure}
\par

Zhou {\it et al.} \cite{Zhou} invoke a Mattheissen's rule justification for extrapolation
to $L \rightarrow \infty$, namely the idea that ``boundary scattering'' causes
$1/\kappa$ to behave like $1/\Lambda + 1/L$.  However, their cell has periodic
boundary conditions and therefore no actual boundary.  The discrete version of the PBE
presented here uses correct statistical theory and incorporates the ring geometry
by design.  Therefore, it should correctly describe the rate at which $\kappa_{\rm eff}(L)$
converges to the $L \rightarrow \infty$ limit, and provide guidance for extrapolation. 
Fig. \ref{fig:power} shows how $\kappa_{\rm eff}(L)$ converges as $L$ increases.
Models with constant mean free path converge exponentially,
as $\exp(-4\Lambda/L)$.  This is true both for continuum and discrete cases. 
This is not surprising, but I have not yet found a simple proof.  The factor of 4
just relates to the fact that only 1/4 of the ring ($\Delta m=N/4$)
is available for decay.  Discrete models with
mean free paths diverging as $1/\omega$ ($p=1$) and $1/\omega^2$ ($p=2$)
are also shown in Fig. \ref{fig:power}.  They apparently converge algebraically,
as $1/L^2$ for $p=1$ and $1/\sqrt{L}$ for $p=2$.  These powers were found numerically
from the slope of the log-log graphs.  I conjecture that the behavior is
$\kappa_{\rm bulk}-\kappa_{\rm eff}(L) \propto (1/L)^{(3-p)/p}$.
This scaling for $p=2$ is in rough
accord with the highest temperature simulation by Zhou {\it et al.} \cite{Zhou},
where convergence was non-linear in $1/L$.  A plot of their numerical results
{\it versus} $1/\sqrt L$ instead of $1/L$ gives a significantly better straight line.
Unfortunately, the extrapolated value then falls below $1/\kappa = 0$, indicating
that even larger simulation cells are needed before extrapolation can be relied on.
Results for GaN by Lindsay {\it et al.} \cite{Lindsay}, using DFT and PBE, agree well with experiment.
They also show that isotopically pure GaN will have $\kappa(300K)\approx 400$W/mK.
This indeed makes $1/\kappa$ closer to 0 than to the extrapolated values of Zhou {\it et al.}.

\section{Discussion}

Non-local heat transport is seldom explicitly \cite{Mahan} discussed.  The non-locality is hidden
if the temperature gradient is uniform.  Then the form of the non-locality (contained in $\kappa(k)$
for a homogeneous system) is irrelevant, since only the $k \rightarrow 0$ limit is seen.
The PBE contains a valid description of non-local response, provided the carrier mean free path
is sufficiently long that a quasiparticle gas description is valid.  
The work described in this paper uses phonon gas theory, as contained
in the PBE, to describe the non-locality.  The PBE is reformulated for
finite size systems with periodic boundary conditions.  
Discretization into parallel slabs is natural for one-dimensional transport.  The resulting
$\kappa(q)$ (Eq.(\ref{eq:kDslabp})) has, I believe, negligible discretization error, 
and correctly includes the nanoscale corrections
that enter when mean free paths are comparable to the 
simulation cell length $L$ in an MD simulation.  This situation is hard to 
avoid for materials with good crystalline order and weak anharmonicity, 
such as the GaN simulation of ref. \onlinecite{Zhou}.

Picturing heat transport as explicitly non-local may have benefits.  For example,
in MD simulations of $\kappa$ by the ``direct method,''
it would be sensible to impose heat in a periodic fashion, ${\dot e}(\ell) = \cos(2\pi\ell/N)$, or
${\dot e}(q)$ containing only the smallest non-zero $q=2\pi/Nd$ allowed.  This simplifies
Eqs.(\ref{eq:dt},\ref{eq:t}).  More important, it should enable extrapolation to the $q \rightarrow 0$
limit more smoothly.  Equally important, it should help reduce the noise level of MD simulations,
because the ``measured'' $T(\ell)$ would be used at all $\ell$ to extract $T(q)$, rather than using
only a few points of $T(\ell)$ to extract a gradient at the midpoint.  A future paper on this
topic is planned.  

A number of recent papers formulate theories of heat conductivity in nanoscale systems
\cite{Lebon, Das, Saaskilahti,Yang}.   Microscopic theory
is needed, not just to supplement simulation,
but more importantly, to aid experiment in interpreting nanoscale effects.
Non-local effects are evident in heat transport by nanoscale samples of good
conductors like graphene.  Landauer methods \cite{Landauer,Angelescu,Mingo} are often preferred, but
have limitations when inelastic scattering is present.  Meir-Wingreen-type \cite{Meir} approaches
often used to supplement Landauer methods in electron transport, and can be generalized
to heat transport \cite{Das}.  The PBE, which is evidently useful to model non-locality, and
includes inelasticity, can perhaps be exploited in new ways to simplify and unify some of these problems. 

\section{acknowledgements}
This work was suggested by discussions with M.-V. Fernandez-Serra and J. Siebert,
whose stimulation is gratefully acknowledged.  I also thank D. Broido, G. Chan, Y. Li, K. K. Likharev, J. Liu,
A. J. H. McGaughey, S. Ocko, T. Sun, R. M. Wentzcovitch, and an anonymous referee for helpful input.
This work was supported in part by DOE grant No. DE-FG02-08ER46550.

\end{document}